# Enhanced plasmonic absorption in spontaneous nanocomplexes of metal nanoparticles with surface modified HPHT nanodiamonds

Vendula Hrnčířová[a]*, Markéta Šlapal Bařinková[a], Muhammad Qamar[a], Kateřina Kolářová[b], Jaroslav Kuliček[a], Štěpán Stehlík[b], Alexander Kromka[b], Bohuslav Rezek[a]

[a]*Faculty of Electrical Engineering, Czech Technical University in Prague, Technická 2, 166 27, Prague, Czech Republic*

[b]*Institute of Physics, Czech Academy of Sciences, Cukrovarnická 10, 162 00, Prague, Czech Republic*

*corresponding author: hrnciven@fel.cvut.cz*

**Abstract**

The combination of nanodiamonds with plasmonic metal particles is being explored for synergic effects that can enhance biosensing and antibacterial treatments, energy harvesting, photocatalysis, and quantum centres. Here we systematically investigate the formation and plasmonic properties of complexes assembled in colloidal mixtures of 20 nm gold or silver nanoparticles with 50 nm HPHT nanodiamonds, the surfaces of which are oxidized or hydrogenated. This approach provides nanocomplexes of well-defined particles with different surface chemistry and electrical conductivity. Nanoscale complexes are formed in each case, as shown by scanning electron microscopy and optical absorption spectra. The stable plasmonic frequency shows that the metal nanoparticles do not become aggregated and keep a few nm gap to nanodiamonds. The concentration dependence study reveals that the plasmonic effect can be significantly increased (up to 124 %) for lower concentrations of metal nanoparticles in the mixtures, whereas their higher concentrations reduce it, regardless of the type of nanodiamond or metal nanoparticle. Based on the experimental results and electromagnetic field-based simulations of the complexes, we discuss a model involving competing charge redistribution and plasmonic interference effects.

**Keywords:** surface modified nanodiamonds, plasmonic nanoparticles, nanocomplexes, colloidal chemistry

1. **Introduction**

Nanoparticles (NPs) have become indispensable for biomedicine and industry, driving the development of nanomaterials with properties tailored for specific applications. Noble metal nanoparticles (mNPs) exhibiting localised surface plasmon resonance (LSPR) are among the most employed. [1] Nowadays, LSPR is commonly utilised in bioimaging, biosensing, thermotherapy [1], energy harvesting [2], and photocatalysis [3], [4]. Plasmon-induced charge transfer at the metal-semiconductor interface has been used to facilitate hot carrier transfer to the semiconductor before thermalisation with a metal phonon bath. [5] LSPR is highly sensitive to changes in surrounding conditions, making it especially useful in biosensors. [1], [6]

LSPR is characterised by the oscillation of free electrons in mNPs (plasmons) resulting in electromagnetic fields near particles such as gold or silver. Oscillations are caused by interaction with light of a specific wavelength that exceeds the particle size. [1], [7], [8] The energy of the photons causes the displacement of the free electrons, inducing a Coulombic restoring force. This results in coherent and collective oscillation leading to an enhanced electromagnetic field around the particle. [1], [4], [6], [7] The plasmonic wavelength is unique to each particle and depends on its type, size, shape, chemical surroundings, and other factors. [1], [4], [6], [7], [8], [9], [10], [11] Despite recent advancements in producing plasmonic particles of various morphologies, establishing accurate and precise plasmonic models remains challenging. [7]

The lack of a detailed mechanistic understanding for plasmon-mediated charge transfer at metal-semiconductor interfaces severely limits the design of efficient application use [5], especially considering that the relaxation of plasmons following particle excitation occurs through complex radiative and non-radiative decay mechanisms. [1], [11] The type of predominant decay is influenced, among others, by particle size. Smaller particles are more likely to undergo non-radiative decay, inducing electric currents, enzyme-like activity, or heat conversion. These effects can be used in antimicrobial applications, or in photothermal therapy and wound healing. On the contrary, larger particles tend to produce a secondary light source widely used in imaging. [1], [4], [6] By adjusting particle parameters, we can optimise optical properties of mNPs including absorption, scattering, and extinction. [1], [7] Such optimisation helps attain the desired performance and enables predicting the plasmonic behaviour for precise applications.

One of the possible alterations of plasmonic properties is the introduction of particles of various kinds. Combining plasmonic particles with other materials can adjust and channel plasmonic resonance or even increase its effect. In many studies, nanodiamonds (NDs) have been found to be great candidates for forming nanocomplexes with plasmonic particles. [9], [10], [12], [13], [14], [15], [16], [17], [18] NDs themselves possess properties convenient for broad modifications. They have also found use in various biomedical practices such as bioimaging, drug delivery, and biosensing. NDs boast biocompatibility, unique optical properties, and versatile surface modifications. [19], [20], [21], [22] The surface of NDs differs from their bulk counterparts, containing a non-diamond carbon phase [23], molecules such as hydrogen or oxygen, and other functional groups. [19], [21], [22] This allows a wide range of surface modifications influencing their stability, conductivity, optical activity, and other properties. Synthesis methods and particle morphology also influence these characteristics. [9], [19], [21], [22], [23], [24] High-pressure high-temperature (HPHT) nanodiamonds have a uniform structure and carbon lattice created from pure carbon precursors with minimum defects. [21], [25], [26] Colour centres such as nitrogen vacancies are often created in HPHT NDs, enabling fluorescence. [9], [19], [20], [27] The monocrystalline structure of HPHT [28] NDs provides a well-defined surface that can be modified through hydrogenation or oxidation leading to opposite effects on certain nanodiamond properties. Hydrogenated HPHT nanodiamonds (HNDs) exhibit surface and electric conductivity, positive zeta potential (ZP), negative electron affinity and a hydrophobic nature [23], [24]. Oxidised HPHT nanodiamonds (ONDs) exhibit surface and electrical resistivity with a negative ZP and possess a positive electron affinity [23] and hydrophilicity [29]. Moreover, they show improved biocompatibility. [27]

By combining diamond and metal nanoparticles, we can create multifunctional systems and leverage their complementary strengths. Nanocomposites of silver and nanodiamonds showed a synergic effect as an antibacterial agent. [12], [17] The effectiveness of plasmonic photocatalysis is also improved by the presence of nanodiamonds. [10] Plasmonic resonance of gold enhanced through the presence of NDs is translated to enhanced photo-responsive reactivity in Surface Enhanced Raman Spectroscopy. [9], [13] NDs channelled heat transfer from plasmonic resonance to their surroundings and so improved the durability of gold particles. [9] Nanodiamonds also improved AuNP photoacoustic signalling, optical imaging, transmission electron microscopy, and catalytic activity. All of this is linked to energy or charge transfer, unique for a hybrid system. [9] This relation is two-way, where the presence of plasmonic particles can also improve the fluorescence of NDs. [1], [15], [16] The multifunctionality of nanocomplexes of plasmonic particles and nanodiamonds opens many possibilities for combining imaging and sensing with therapeutic applications such as thermotherapy [9], [14], [18]. The ultimate properties of these nanocomplexes depend on many factors such as the type of constituents, their ratio, size, distribution, shape, aggregation, chemical surroundings, and mutual interaction. [1], [6], [9]

Nanocomplexes are often formed by direct synthesis of mNPs on the ND surface by reducing metal salts or ions using reducing and stabilizing agents. [9], [10], [26], [28] Although the particle size distribution can be regulated via synthesis duration, temperature and pH [9], this approach faces some challenges. Control over the shape, size, and distribution of mNPs on ND surfaces remains limited. [9] Consequently, achieving consistent plasmonic resonance and reproducibility in the nanocomplexes is difficult. The post-synthesis attachment of well-defined mNPs on ND surfaces through chemical bonds can partially overcome these difficulties. However, the system of post-synthesis nanocomplexes is more challenging in terms of chemical control, toxicity and stability. [9] Even though many studies observed unique advantages of the partnership between plasmonic particles and nanodiamonds, the mechanisms are still not fully explained, and the resulting properties are not securely predictable. [9] Consistent and well-defined formation of nanocomplexes and systematic characterization are essential to elucidate the fundamental plasmonic mechanisms.

In this study, we investigate interactions in colloidal mixtures of mNPs and NDs, combining experimental data with COMSOL Multiphysics simulations. The use of colloidal mixtures ensures well-defined particle characteristics, which promotes consistency in the formation of nanocomplexes and their properties. We explore the influence of metal nanoparticle concentration on plasmonic absorption of the mixtures and incorporate both hydrogenated and oxidized HPHT nanodiamonds that possess contrasting properties to analyse a broader range of variables. We show that reproducible mNP-ND nanocomplex formation can occur spontaneously, without need of complicated linker strategies. We analyse how interactions between nanoparticles influence the plasmonic properties of mNPs, leading to significantly enhanced plasmonic absorption due to ND presence. We discuss potential mechanisms behind these effects and propose a comprehensive model that may lay a foundation for novel design principles of nanodiamond-based plasmonic nanocomplexes, especially, but not only for biomedical applications.

## 2. Materials and Methods

### 2.1. *Materials*

*Surface modified HPHT NDs*

A commercially available HPHT monocrystalline synthetic nanodiamond powder (MSY 0-100nm, Pureon) with a particle size range of 0 – 100 nm and median size of 50 nm was used for surface modification. Oxidized nanodiamonds (ONDs) were obtained by oxidizing the nanodiamond powder in an oven at 450°C for 5 hours. [26] To prepare hydrogenated nanodiamonds (HNDs), the nanodiamond powder was activated for 6 hours in a hydrogen atmosphere at 800°C [28]. Colloidal suspensions were obtained by mixing 1 mg of each type of ND powder with 1 mL of demineralized water and sonicating for 1 hour using an ultrasound tip at 200 W for proper dispersion (Hielscher, UP200s). [28] Likewise, before each experiment, the colloidal nanodiamonds were sonicated at 35 kHz for 10 minutes in a sonication bath (Bandelin, Sonorex Digitec DT 31 Ultrasonic Bath) to resuspend potential sedimented particles.

*Metal nanoparticles*

We used colloidal gold (GC20, BBI Solution) of a 50 µg/mL stock concentration with a spherical shape and particle size of 20 nm (AuNPs) containing sodium citrate as a capping agent. We employed silver spherical nanoparticles in colloid (730793, Sigma) of a concentration of 20 µg/mL and particle size 20 nm (AgNPs) stabilized by sodium citrate. Both colloids were stored at 4 °C in the dark place.

*Colloidal mixtures of NDs and mNPs*

A four-step two-fold dilution in HPLC water (Roth) of the ND stock solution of 1 mg/mL was performed resulting in a concentration of 250 or 62.5 µg/mL of NDs, taken from the second and fourth step, respectively. A four-step two-fold dilution series of AgNPs and AuNPs was conducted for each type of mNP (20, 10, 5, 2.5 µg/mL or 50, 25, 12.5, 6.25 µg/mL, respectively). From these suspensions, a total of 32 mixtures was prepared by mixing the ND colloids with the NP colloid in a 1:9 ratio. The resulting samples were: AuHND25, AuHND6, AuOND25, AuOND6, AgHND25, AgHND6, AgOND25, and AgOND6, where numbers denote the final concentrations of NDs in µg/mL. Each mixture type further contains 4 samples of different mNP concentrations. Mixtures were agitated by hand. **Figure 1** (A) shows the scheme of colloidal mixture preparation and resulting sample sets. A new set of colloidal mixtures was prepared for each experiment to ensure well-defined comparable results. Figure 1 (B) presents a photo of a well-plate containing pristine colloids and colloidal mixtures as used for further spectroscopy analysis. All the samples exhibit high optical transparency.

### 2.2. *Analyses*

*UV-Vis spectroscopy*

UV-Vis spectroscopy of colloidal mixtures was measured using the Epoch2 microplate reader (BioTek) with a xenon lamp and a monochromator wavelength selection from 200 to 900 nm with a 2 nm increment. Solutions were analysed in triplicate in a 96-well UV black body microplate without a lid. Samples were measured at 37 °C and underwent a 10-second double orbital shake at 282 cpm before measuring. Individual colloids of mNPs and NDs were measured in corresponding concentrations as references. HPLC grade water was used as a blank sample. The stability of colloids was measured by comparing the absorption spectra of samples at time zero and after 24 hours. Between measurements, the samples were stored with a lid in an incubator (Memmert, UNB 400) at 37 °C. Absorption spectra were analysed using MATLAB (MathWorks, R2023a) [30].

*Dynamic light scattering and zeta potential*

Dynamic light scattering (DLS) and zeta potential (ZP) were measured by Zetasizer Nano (Malvern Panalytical) with a helium-neon laser (633 nm) and scattering angle of 173°. Samples were measured in a disposable folded capillary cell equipped with electrodes. Before measuring, the samples underwent a 1-minute stabilization period. Acquired results were inspected using the quality report from Zetasizer to ensure the reliability of the measurements. Measurements were performed in triplicate. The pH of each sample was measured using the Malvern MPT-2 Multipurpose Autotitrator at room temperature after the zeta potential measurements.

*Scanning electron microscopy (SEM)*

Scanning electron microscopy (SEM) was performed using the Zeiss EVO 10 system at 17 kV acceleration voltage, 5 pA probe current, and approximately 8.5 mm working distance. SEM micrographs were acquired simultaneously using the signal from back-scattered electrons (BSE) and secondary electrons (SE). Samples were drop-cast onto a silicon wafer substrate and left to air-dry at room temperature. To enhance image representation, a mask was created using the BSE signal by thresholding, thus distinguishing mNPs from NDs. This mask was coloured red and overlayed over the SE image using Gwyddion Software [31] and MATLAB (MathWorks, R2023a) [30].

*Transmission electron microscopy (TEM)*

Transmission electron microscopy (TEM) was performed using the Tecnai G2 20 (FEI) TEM with a LaB6 cathode at an acceleration voltage of 200 kV and an Olympus Veleta CCD camera. The TEM samples were prepared by immersing a carbon-coated copper grid (SPI Supplies) into the nanocomplex colloid and allowing it to dry before imaging.

2.3. *Numerical Simulations*

COMSOL Multiphysics software with the RF Module was used for the modelling of the mNP-ND nanoparticle complexes geometry and for depicting their interactions with an incident electromagnetic field. The spherical diamond particles with a 50 nm diameter were placed at the centre of the nanocomplex. Diamond material properties were set according to standard bulk values. [32] Diamond electrical conductance was adjusted to simulate effect of hydrogenated HPHT nanodiamonds exhibiting surface conductivity ($10^{-5}$ S/cm) [24] or highly electrically resistive oxidized HPHT nanodiamonds ($10^{-15}$ S/cm) [33] that were employed in the experiments. Gold and silver spherical particles with a 20 nm diameter were distributed symmetrically around the central ND. The separation between the metal particles and distance to the nanodiamond was varied in a range of 0-10 nm to analyse the plasmonic effects according to nanoparticle distance and contact. The dielectric properties of metal nanoparticles were set according to the data given by the Johnsons and Christy model [34]. Their electrical conductance was set according to table values. [33] The background electric field was set at 1 V/m. Perfectly matched layers (PML) around the model were set at a five times larger diameter than the nanodiamond size in order to eliminate background reflection from the surroundings and to absorb outgoing waves. The finite element method (FEM) was used to solve Maxwell's equations of the mNP-ND models. A fine mesh setting and mesh refinement was employed for a sufficiently detailed electric field analysis around the nanoparticles.

3. **Results and Discussion**

   3.1. *Optical Absorption of Colloidal Mixtures*

The experimental UV-Vis absorption spectra of pristine metal nanoparticles and nanodiamond colloids of various concentrations as used for the mixtures are plotted in Figure 1 (C). The spectra of both types of NDs are featureless and are similar to those of hydrogenated and oxidised nanodiamonds reported by other groups. [35] They exhibit low absorption in the visible region, where the presence of colour centres is not manifested [36], and a strong absorption rise in the UV region due to the diamond bandgap of 5.5 eV. Rayleigh scattering can also enhance UV absorption in particles significantly smaller than the incident light. The UV-Vis absorption spectra of the metal nanoparticles displayed typical plasmonic resonance absorption peaks. Silver (AgNPs) and gold (AuNPs) nanoparticles had absorption peaks indicating plasmonic resonance around 396 nm and 524 nm, respectively. The plasmonic peak positions and linewidths were stable across all investigated concentrations. A change in the trend of the absorption spectra of mNPs in the UV region, starting around 220 nm, is a characteristic feature of the interband transitions in AgNPs and AuNPs.

In addition to increased optical absorption due to LSPR, we also observed an enhanced electromagnetic field of the particles due to plasmonic resonance. Figure 1 (D, E, F) shows the electromagnetic field of particles at 522 nm obtained from the COMSOL Multiphysics RF Module. The electromagnetic field of a single diamond particle was lower compared to that of the gold nanoparticle. The electromagnetic field of a single AuNP was symmetrical and polarized with a maximum observed at a wavelength of 522 nm. This wavelength corresponded to the plasmonic resonance wavelength we observed during our experiments. In the nanocomplex (Figure 1 (F)), we saw the electromagnetic field concentrated mostly between the gold and diamond particles. The gold electromagnetic field remained polarized but lost its symmetry. Moreover, the intensity of the nanocomplex electromagnetic field reached 2-times higher values than the intensity of the AuNP alone.

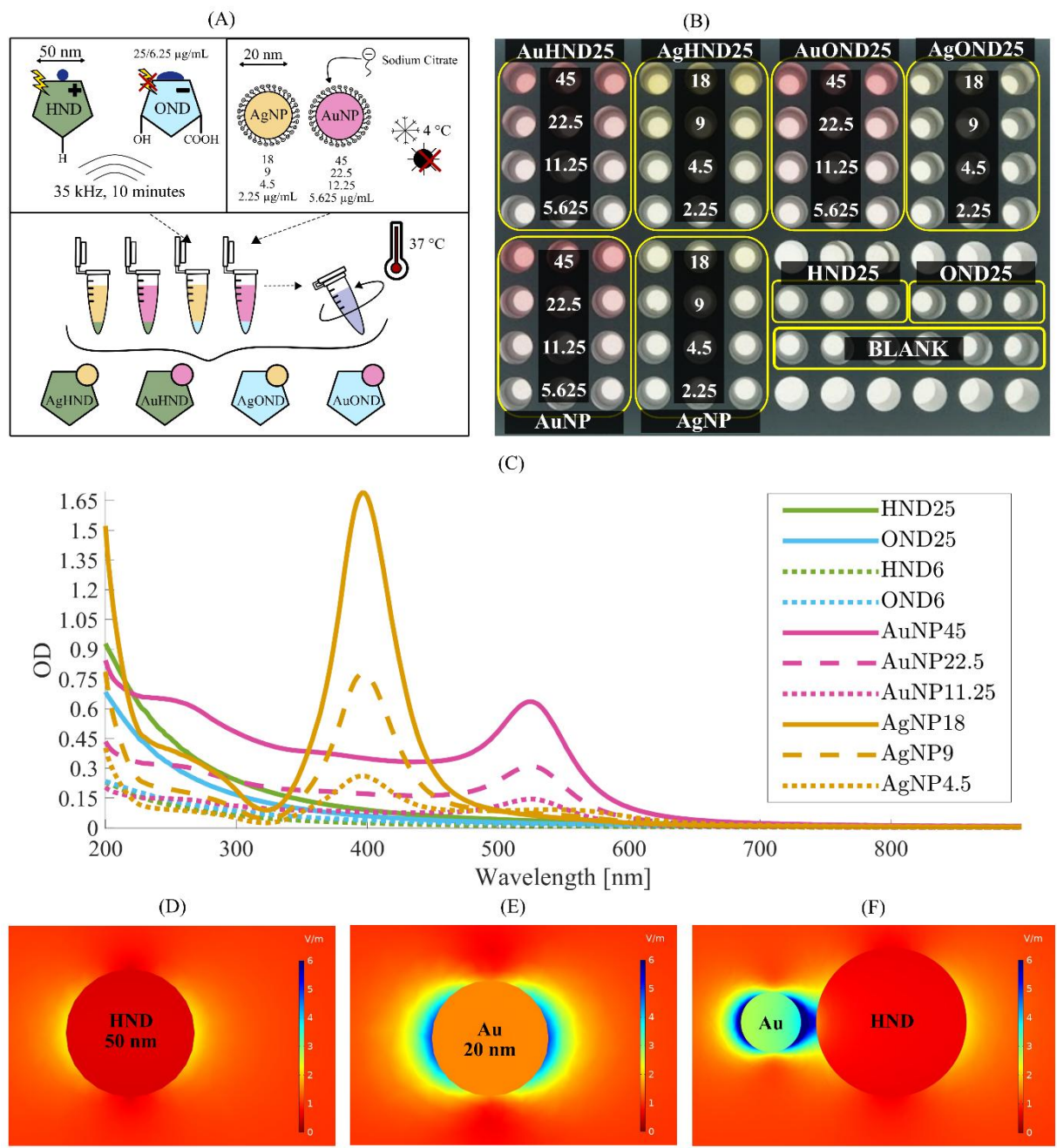

*Figure 1: (A) Scheme of the employed colloidal nanoparticles and colloidal mixture preparation. (B) A photo of the well-plate containing pristine colloids and colloidal mixtures. The numbers indicate the concentration of metal nanoparticles in the sample, expressed in µg/mL. (C) Optical UV-vis absorption spectra of nanodiamonds and metal nanoparticles in colloidal solutions of different concentrations as obtained from experimental measurements. The number next to a nanoparticle name represents the actual concentration of that nanoparticle in the colloidal solution. Images below are the maps of electromagnetic field intensity computed by COMSOL RF Module simulations of (D) 50 nm conductive nanodiamond, (E) 20 nm gold nanoparticle, and (F) nanocomplex of 20 nm gold nanoparticle and 50 nm nanodiamond with a gap of 5 nm, all in water media. In all images the electric field intensity is mapped at 522 nm wavelength, corresponding to plasmonic absorption peak of the gold nanoparticles.*

**Figure 2** shows optical absorption spectra of colloidal mixtures with different concentrations of metal nanoparticles and a fixed concentration of surface modified nanodiamonds. The absorption spectra clearly show the presence of both types of nanoparticles in the colloidal mixtures, with the high UV absorption inherited from NDs and a plasmonic peak originating from mNPs. All prepared colloidal mixtures followed the Lambert-Beer law measured at 450 nm, outside plasmonic peaks. As the mNP concentration decreases, the spectra keep their shape and position of the plasmonic peak, only the overall absorption (optical density, OD) of the colloids decreases. No specific new features are obvious from the spectra. Thus, the question arises whether some interaction between the mNPs and NDs is present in the mixtures. Therefore, we performed structural analysis and subsequently analysed the plasmonic peaks in detail.

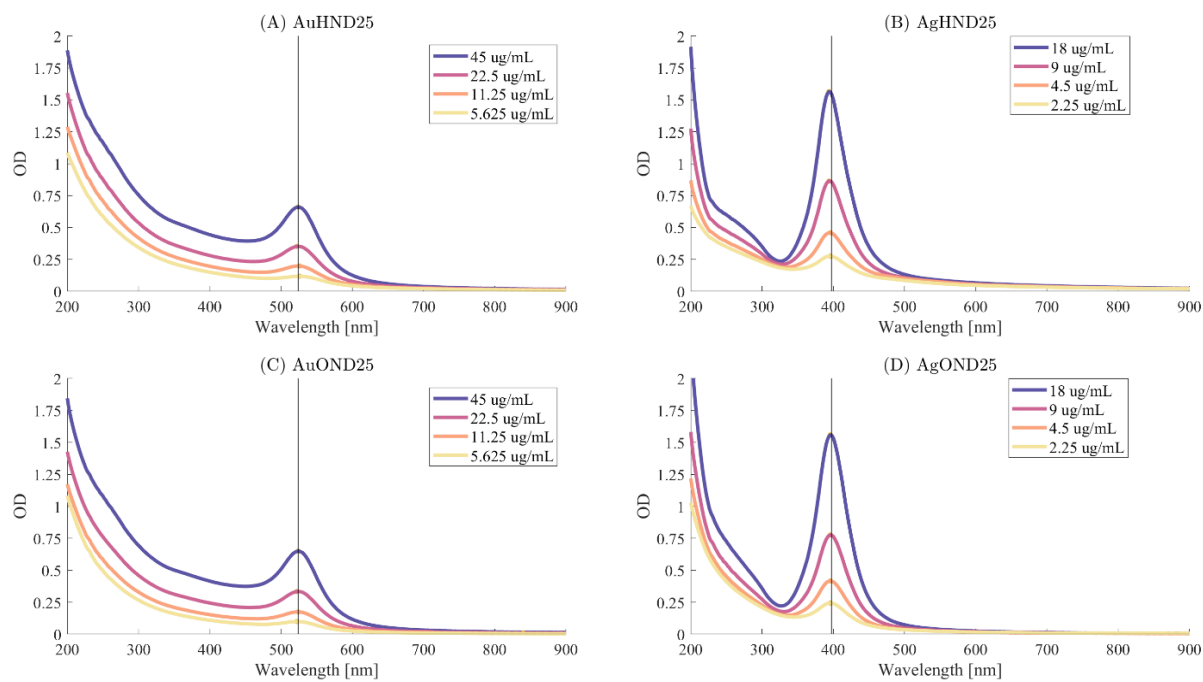

*Figure 2: Optical absorption spectra of colloidal mixtures with different concentrations (as shown in the legend of each graph) of metal nanoparticles and with fixed 25 µg/mL concentration of surface modified 50 nm nanodiamonds: (A) AuHND25, (B) AgHND25, (C) AuOND25, and (D) AgOND25. The data curves are averages from triplicate, blank-corrected samples.*

3.2. *Nanocomplex formation and structure*

**Figure 3** (A, B, C, D) shows composite SEM images of typical colloidal mixtures (AgHND25, AgOND25, AuHND25, and AuOND25) at the same magnification. The nanoparticle concentrations in these mixtures were ND 25 µg/mL, AgNP 9 µg/mL and AuNP 22.5 µg/mL. The morphology of particles is represented by the SE signal and red-marked areas corresponding to metal nanoparticles are labelled based on the BSE signal.

According to the images, the red-marked mNPs are localised on ND surfaces (grey) and do not appear elsewhere. The mNPs are consistently distributed across nanodiamond surfaces, particularly in the OND mixtures. Therefore, they must have been on the ND surface before the drop-casting. This is the first indication that mNP-ND nanocomplexes can be spontaneously formed in the mixtures. It is noteworthy that the nanocomplex formation seems to occur irrespective of ND or mNP type.

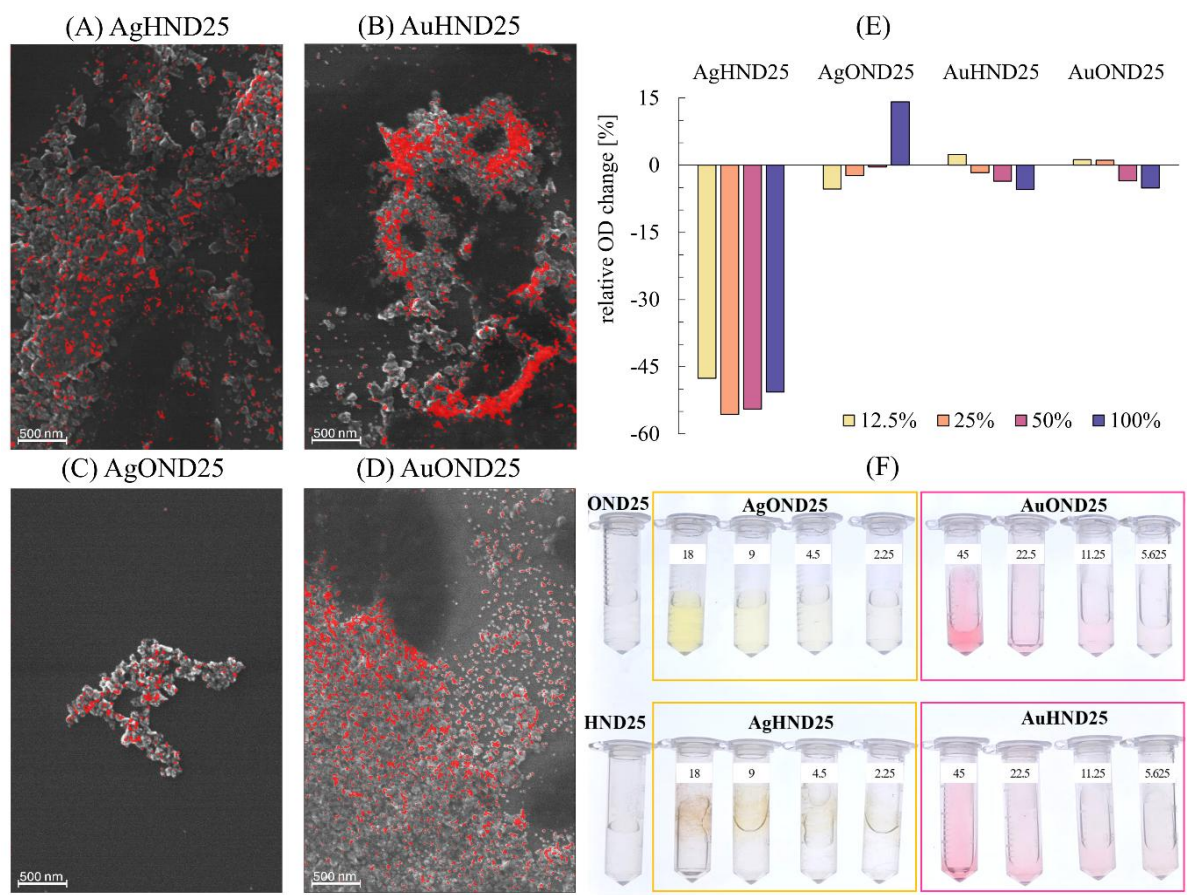

*Figure 3: Composite SEM images of colloidal mixtures - (A) AgHND25, (B) AuHND25, (C) AgOND25, and (D) AuOND25. The nanoparticle concentrations in these mixtures were ND 25 µg/mL, AgNP 9 µg/mL and AuNP 22.5 µg/mL. The morphology of particles is represented by the SE signal and red-marked areas corresponding to metal nanoparticles are labelled based on the BSE signal. (E) Relative optical density changes of colloidal mixture absorption peak observed after 24 hours. 100% bars (purple) correspond to 18 µg/mL for silver particles and 45 µg/mL for gold particles in the mixture. The zero line corresponds to the OD of mixtures at time zero. (F) A photo of colloidal mixtures in Eppendorf tubes after 24 hours. Values on the tubes are concentrations of metal nanoparticles in the mixture in µg/mL.*

To elucidate the possible nanocomplex formation, we performed DLS analysis of the pristine and mixed colloids. **Table 1** summarizes the results of the DLS analysis of the employed colloidal nanoparticles and resulting nanocomplexes: mean size (from the number distribution), pH, and zeta potential (ZP). The number distributions of all samples are provided in the Supplementary material in Figure S1. In DLS, all the mNPs exhibit sizes slightly smaller (16 nm) than the nominal 20 nm, which shows possible limitations of the measurement method. Both mNP types have negative ZPs of -36 mV and -41 mV, in agreement with their standard properties and good colloidal stability. AgNPs (2.25 µg/mL) exhibit a basic pH of 7.1, while AuNPs (5.625 µg/mL) show an acidic pH of 6.5. HNDs exhibit sizes larger (79 nm) than the nominal 50 nm, a positive ZP of +21 mV, and an acidic pH of 6.33, also in agreement with the commonly reported properties of hydrogenated nanodiamonds. However, their mixtures, AgHND and AuHND, exhibit negative ZPs. With increasing mNP concentrations in the mixture, the ZP becomes larger (more negative), from -16 mV to -34 mV and from -17 mV to -24 mV, respectively.

The weight (as well as number) concentration of HNDs in the measured mixtures is up to 10× higher than that of mNPs. Thus, it is unlikely that the ZP of mNPs would prevail in the measurements if the particles acted separately in the mixtures. The effect on ZP is similar to the formation of the protein corona on nanodiamonds. [37], [38] Although not consistently manifested in the DLS size measurements (which have some principal limitations in general), the change of ZP polarity from positive to negative is the second indication that a nanocomplex is formed from HNDs and mNPs, corroborating the SE/BSE data in Figure 3 and the TEM images in Figure 7. AgHND25 DLS measurements report mean sizes on a micrometre scale indicating large agglomerates in the mixtures. This finding is further discussed below, especially in the context of the AuHND25 mixture showing particle mean sizes from 24 nm to 79 nm based on the mNP concentration.

ONDs also exhibit a size larger (68 nm) than 50 nm, a negative zeta potential of -40 mV, and an acidic pH of 6.12, also according to expectations. The ZP of AgOND25 mixtures remained roughly stable across all mNP concentrations. For AuOND25, ZP values become smaller (less negative) with increasing mNP concentration (from -40 mV to -24 mV), again in similarity to protein corona formation. [37], [38] Although ZP value changes of ONDs may not be fully conclusive, considering the SE/BSE data in Figure 3 and TEM data in Figure 7, they corroborate nanocomplex formation of mNPs with ONDs, too. DLS measurements show variations in the mean sizes depending on mNP concentrations in the mixtures. At lower mNP concentrations, the mean sizes of both OND mixtures (AgOND25 and AuOND25) are close to the OND size, whereas at higher concentrations, the mean sizes align with those of the mNPs. This can be attributed to the saturation of the complexes or the increase in pH, which alters the colloidal environment conditions. All OND samples have basic pH ranging increasing with mNP concentration from pH 7 to 7.36 except for AuOND25 5.625 µg/mL which is acidic with a pH of 6.77.

For the particles to interact, they had to come into proximity of each other. Positively charged HNDs can attract with the negatively charged mNPs electrostatically and thus assemble into nanocomplexes. However, the negative ZP of ONDs implies electrostatic repulsion with mNPs. Therefore, another force must overcome this repulsion. We assume that a chemical reaction can happen between the sodium citrate on the mNPs and oxygen-containing groups on the ONDs surface. Cross-linking with citric acid through covalent intermolecular ester interactions with –COO and –OH groups has been widely established in the synthesis of biopolymers. [39]

We also studied the colloidal stability of mixtures after 24 hours. Figure 3 (E) shows that all the mixtures were reasonably stable, with OD variations within ±10%. The only exception was AgHND25, which showed a consistent optical absorption decrease after 24h by around 50% in all concentrations. Its ZP value is higher than that of the AuHND25 mixture (Table 1), which is considered stable. It is important to remember, however, that ZP is only one of the factors for colloidal stability. Thus, the instability is most likely due to agglomeration specific to AgHND25, as indicated also in the SEM image in Figure 3 (A). Additionally, we observed dark clusters in AgHND25 samples on the walls of the Eppendorf tube after 24 hours in Figure 3 (F). The residual of the AgHND25 solutions appeared without colour. In contrast, other samples retained their original colour, and no visible sedimentation was present. The agglomeration is further supported by the large particle sizes detected in DLS measurements (Table 1).

Despite the presence of clusters, AgNPs must be separated on the ND surfaces as the plasmonic peak position remains stable. This raises the question of what is behind the instability of the AgHND25 sample. One possibility is that AuNPs on the HND surface are more effective at preventing agglomeration of unstable HND than AgNPs. As shown in Table 1, AgHND25 samples have a more basic pH, while AuHND25 are more acidic. The presence of mNPs can alter the colloidal environment, which may influence the stability of the HNDs. Additionally, in the AgHND25 sample, fewer nanocomplexes may be formed, allowing more isolated HND to agglomerate together, increasing mean size and changing absorption. Nevertheless, AgHND25 exhibits significant changes in plasmonic peak intensity compared to pristine AgNP colloid compared to other nanocomplexes showing strong interactions between AgNPs and NDs (Figure 5). Note that the stability analysis was executed after 24h, thus it does not affect other experiments presented here, which were always performed with freshly made solutions and mixtures.

*Table 1: Results of DLS analysis of the employed colloidal nanoparticles and resulting nanocomplexes: mean size (from number distribution), pH, and zeta potential. Concentrations for colloidal mixtures denote concentration of mNPs in the mixtures with the fixed ND concentration of 25 µg/mL.*

| Sample | Concentration [µg/mL] | Mean size [nm] | pH | ZP [mV] |
|---|---|---|---|---|
| HNDs | 25 | 79 | 6.3 | +21 |
| ONDs | 25 | 68 | 6.1 | -40 |
| AgNPs | 2.25 | 16 | 7.1 | -36 |
| AuNPs | 5.625 | 15 | 6.5 | -41 |
| AgHND25 | 18 | 1850 | 7.3 | -34 |
|  | 9 | 220 | 7.3 | -28 |
|  | 4.5 | 3091 | 7.1 | -24 |
|  | 2.25 | 4801 | 7 | -16 |
| AgOND25 | 18 | 21 | 7.4 | -41 |
|  | 9 | 21 | 7.1 | -40 |
|  | 4.5 | 59 | 7 | -39 |
|  | 2.25 | 68 | 7 | -40 |
| AuHND25 | 45 | 26 | 7 | -24 |
|  | 22.5 | 24 | 6.8 | -19 |
|  | 11.25 | 24 | 6.6 | -21 |
|  | 5.625 | 79 | 6.5 | -17 |
| AuOND25 | 45 | 18 | 7.1 | -24 |
|  | 22.5 | 21 | 7.1 | -31 |
|  | 11.25 | 68 | 7 | -35 |
|  | 5.625 | 73 | 6.8 | -40 |

3.3. *Plasmonic effect in nanocomplex*

The above structural analysis indicated the formation of mNP-ND complexes in all composite combinations. This may also affect plasmonic interactions. Such effects were not obvious from the broad range spectra in Figure 2. Thus, we performed a detailed analysis of the plasmonic peaks themselves. **Figure 4** shows detailed plasmonic absorption spectra of colloidal mixtures with fixed ND contractions of 25 µg/mL or 6.25 µg/mL and varying concentrations of mNPs. In theory, if the particles did not interact with each other, their spectra would add up in the colloidal mixtures. Thus, we subtracted the optical absorption spectra of NDs from the spectra of their mixtures (dashed lines) resulting in the subtracted spectra (full lines). One can immediately notice that the subtracted spectra do not align with the spectra of pristine mNPs of corresponding concentrations (dotted line). One can see both enhancement and suppression of plasmonic absorption intensities. When there is a lower concentration of mNPs, the presence of NDs causes an enhancement in optical absorption. As the concentration of mNPs increases, this enhancement decreases. Eventually, we observed the suppression of optical absorption at a high concentration of mNPs. Surprisingly, this declining trend appeared in all our samples and was independent of the ND or mNP type used in the mixtures. We evaluated the peak intensities to show the trends more clearly.

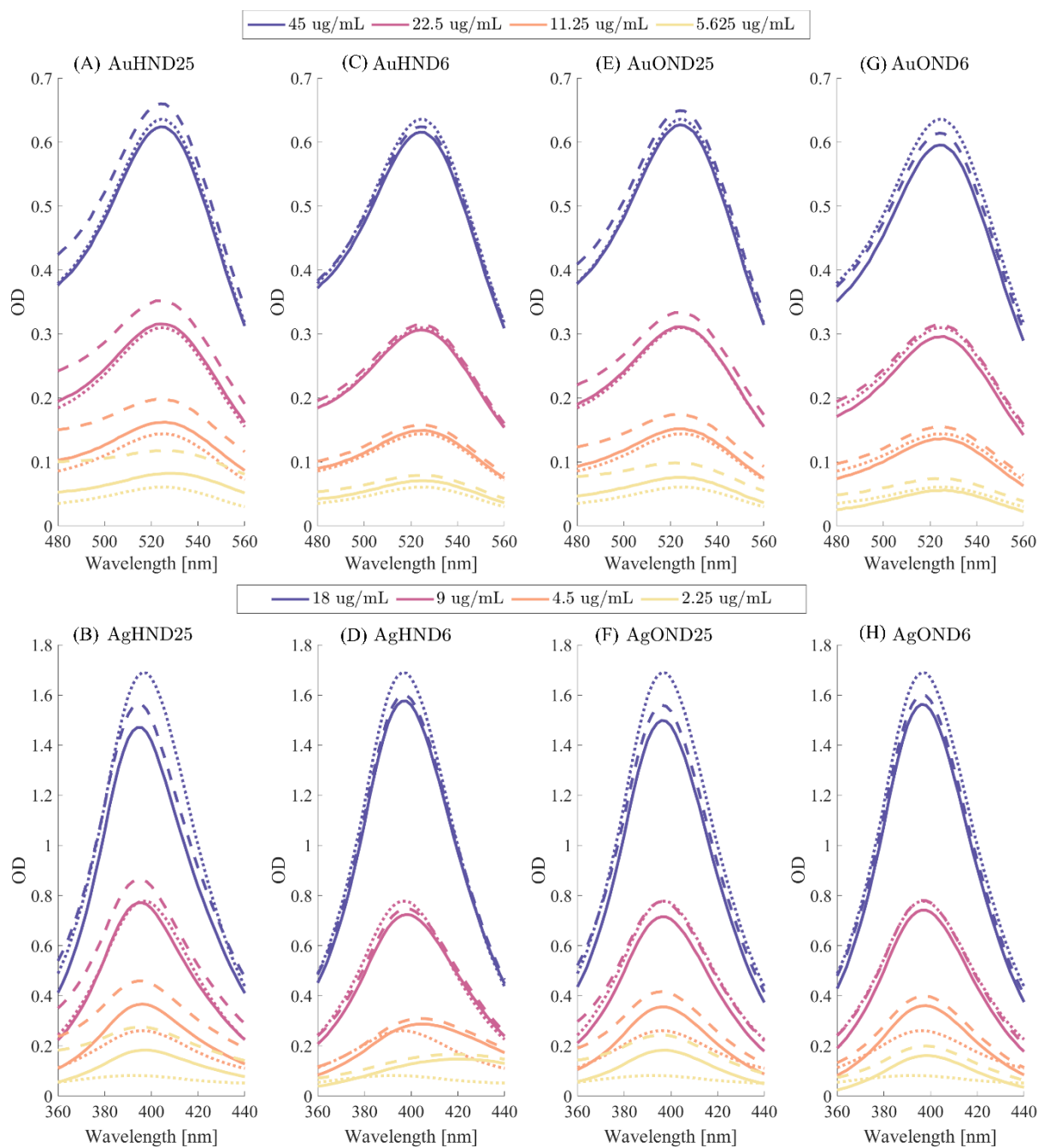

***Figure 4:*** *Detailed plasmonic absorption spectra of colloidal mixtures with 25 µg/mL of NDs: (A) AuHND25, (B) AgHND25, (E) AuOND25, (F) AgOND25 and spectra of mixtures with 6.25 µg/mL of NDs: (C) AuHND6, (D) AgHND6, (G) AuOND6, (H) AgOND6. The measured spectra are represented by the dashed lines. Full lines denote the spectra after subtraction of the respective pristine nanodiamond spectra. Dotted lines represent pristine mNP spectra. The values in the legend denote concentration of metal nanoparticles in the mixtures.*

**Figure 5** summarizes relative enhancements and suppressions of plasmonic peak absorption intensity (in terms of OD) evaluated for different mNP concentrations in the colloidal mixtures. The zero line corresponds to the plasmonic peak OD of the pristine mNP colloids at the respective concentration. The mNP concentration of 100% (purple) corresponds to 18 µg/mL of silver nanoparticles and 45 µg/mL of gold nanoparticles. Analysis also shows influence of ND concentration and surface treatment (hydrogenated or oxidized). Concentrations of NDs in mixtures were 25 µg/mL and 6.25 µg/mL and can be found in the mixture names.

Silver nanoparticles were significantly more influenced than gold nanoparticles by the presence of NDs in the mixtures. AgHND25 and AgOND25 of the lowest concentration (12.5%, i.e. 2.25 µg/mL) exhibit a more than doubled intensity of the plasmonic peak, 124 % and 123 %, respectively, compared to the pristine AgNP colloid of the same concentration. AgHND25 and AgOND25 with the highest mNP concentration (100%, i.e. 18 µg/mL) also had the largest decrease of OD intensity by -13% and -11%, respectively.

The presence of NDs caused only around 35% and 25% increase in the plasmonic intensity of AuHND25 and AuOND25 at the lowest gold nanoparticle concentration (12.5%, i.e. 5.625 µg/mL). At high gold concentrations in the mixture (100%, i.e. 45 µg/mL), we observed around a 1.5% decrease in both gold nanocomplex mixtures.

As for ND concentration influence, the lowered concentration of NDs in the mixture (6.25 µg/mL) caused a smaller enhancement effect compared to the standard concentration (25 µg/mL). In the case of the AuOND6 mixture, the data showed solely a decrease reaching from -4.5% to -8.2% for all concentrations. Overall, there is an obvious declining trend in the plasmonic effect with increasing mNP concentrations in nanocomplexes, irrespective of metal or nanodiamond type.

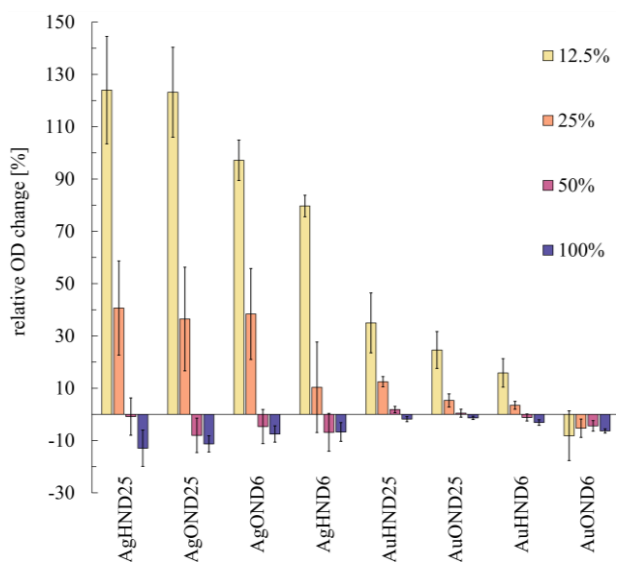

*Figure 5: Relative gains and losses in plasmonic peak absorption intensity (OD) measured in the colloidal mixtures as a function on mNPs concentration (100% (purple) corresponds to 18 µg/mL of silver nanoparticles and to 45 µg/mL of gold nanoparticles) and hydrogenated or oxidized NDs concentration (25 and 6.25 µg/mL) in the mixture. The zero line corresponds to plasmonic peak OD of the pristine mNP colloids (at 398 nm for AgNPs and 524 nm for AuNPs) at the respective mNP concentration. The results are complemented by error bars computed from the combined uncertainty of the standard error of the means.*

We conducted theoretical simulations to help explain the results of our experiments. The plasmonic model consisted of mNPs of varying numbers from 2 to 14 with or without a ND particle in their centre. **Figure 6** (A) presents an example of such nanocomplex geometry. Models with conductive ($10^{-5}$ S/cm) and non-conductive ($10^{-15}$ S/cm) NDs represented mixtures with HNDs and ONDs, respectively.

However, we did not observe significant differences in the plasmonic resonance between conductive and non-conductive nanodiamonds used in the model. Figure 6 (B) plots the differences in the respective plasmonic absorption peak intensities of the nanocomplex with and without a nanodiamond in the centre, depending on the number of metal nanoparticles (i.e. on different mNP concentration). In other words, it shows the effect that the presence and conductivity of nanodiamonds may have on plasmonic absorption with an increasing concentration of mNPs in the nanocomplex. With an increasing number of up to 6 mNPs in the cluster with ND, the absorption gain decreases, which corresponds to the experiments. From 8 to 14 particles, we observed a saturation or even oscillatory trend, which could be related to the interference of the electromagnetic field among mNPs.

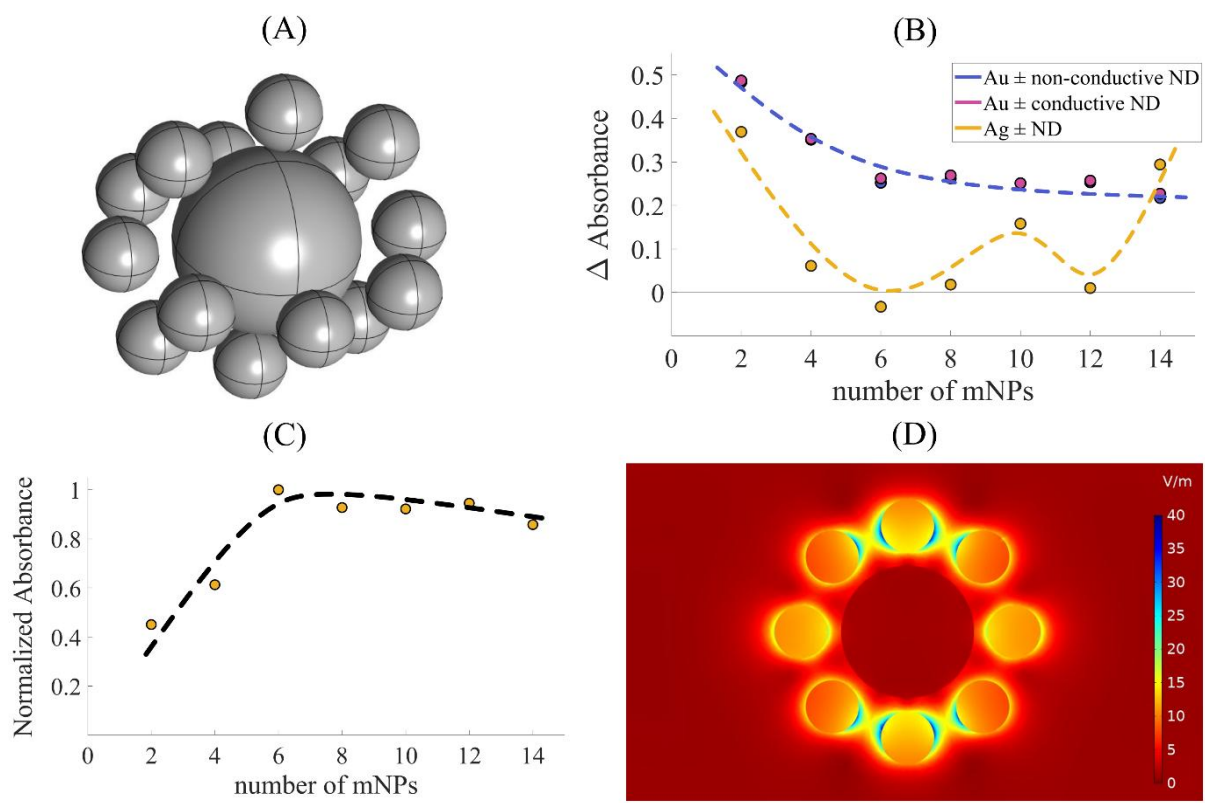

*Figure 6: Results of nanocomplex simulations using COMSOL Multiphysics RF Module: (A) General geometry of the nanocomplex model, in this case with 14 mNPs (20 nm diameter) around the central ND (50 nm diameter) and with gaps at least 5 nm between the nanoparticles. (B) Difference in plasmonic absorption (at the corresponding plasmonic wavelength for Au or Ag nanoparticles) between plasmonic models with and without ND (denoted as ±) in the centre for varying numbers of mNPs (i.e. different mNP concentration) in the nanocomplex and subtracted background. The dashed lines are a guide for the eye. (C) The trend in plasmonic peak intensity in the model with various numbers of AgNPs in the nanocomplex without the presence of nanodiamond in the centre. The dashed line is a guide for the eye. (D) 2D map of the electromagnetic field intensity at the AgNP plasmonic wavelength computed for the nanocomplex with 14 silver nanoparticles around a conductive nanodiamond in water as surrounding medium.*

Figure 6 (C) shows the plasmonic absorption of the model consisting of only silver particles of varying numbers, i.e. corresponding to their higher concentration. Adding up to 6 silver particles caused an increase in plasmonic absorption. However, adding more than 6 particles to the model leads to saturation and a slight decrease in the plasmonic absorption. This shows that there is some interaction that quenches further plasmonic oscillations.

In addition, a significant plasmonic peak shift towards larger wavelengths occurred in models containing 12 and 14 particles. This indicates that the nanoparticles act as aggregates in those highly concentrated systems. We did not observe this behaviour experimentally in the colloids of pristine mNPs or in mNP-ND complexes. It confirms that the metal particles remain separated enough from each other due to mutual electrostatic repulsion.

However, the saturation of plasmonic absorption in Figure 6 (C) occurs from 8 mNP per cluster. As illustrated by Figure 6 (D), we observed an uneven distribution of the electromagnetic field in models with a higher number of metal nanoparticles. This occurred regardless of the presence of a nanodiamond in the centre of the system. Thus, it must be electromagnetic interference that quenches the plasmons intensity in such systems.

### 3.4. *Plasmonic wavelength*

Despite changes in intensity of the measured UV-Vis spectra, the wavelength of the plasmonic absorption peak (often referred to as plasmonic frequency) of the nanocomplexes did not change compared to the peaks of pristine colloids. All mixtures independent of mNP concentrations or type of nanodiamonds showed stable peak wavelengths as seen in Figure 2. Generally, a plasmonic peak shift indicates a size change of particles due to aggregation or coating by other materials in core-shell structures. [5] A shift to larger wavelengths (lower frequencies) occurs with increasing sizes. [6], [40] Thus despite the clear effect in plasmonic intensity, the experimental results seem to indicate that there is no such close interaction among the mNPs or between mNPs and nanodiamond. Simulations were thus performed to elucidate this issue.

**Figure 7** (A) shows the plasmonic absorption peak position of mNP-ND and mNP-mNP dimers in water as evaluated from electromagnetic field simulations using the COMSOL Multiphysics RF Module. The distance between the particles is varied from 5 nm to direct contact (0 nm). For two metal nanoparticles in contact, the plasmonic resonance red-shifts by 100 nm and 80 nm for AgNPs and AuNPs, respectively. Accompanying the decreasing particle distance, there was also an increase in the intensity of the electromagnetic field. The large shift is due to the two metal nanoparticles in contact behaving as one larger particle, in agreement with standard theory. However, the shift occurred even for a small gap between the particles, e.g. at a 1 nm gap, the shift was still 63 nm and 45 nm for AgNPs and AuNPs, respectively. The simulations show that mNPs must be > 5 nm apart in the nanocomplex to prevent a wavelength shift. This distance is in agreement with the exponential distance decay of interparticle plasmon coupling. [41] Similarly, when mNPs were in direct contact with the nanodiamond, the plasmonic wavelength shifted by 10 nm for AuNP and 20 nm for AgNP from their original plasmonic wavelengths. The shift was negligible once of the gap size increased to 3 nm. The simulations thus clearly show that the mNPs and nanodiamond are at least 3 nm apart in the nanocomplexes and not in direct contact.

Figure 7 (B) shows the geometrical scheme of the mNP-ND nanocomplexes depicting the minimum necessary gaps to prevent plasmonic wavelength shifts. Sodium citrate coating and mutual electrostatic repulsion can explain the gaps between the mNPs when they spontaneously assemble on nanodiamonds, thereby preventing the plasmonic shift. The 3 nm gap between mNPs and NDs also correlates reasonably with the citrate coating and its molecular links with the diamond surface chemical groups.

Figure 7 (C-F) shows transmission electron microscopy (TEM) images of the nanocomplexes. We observed individual particles of both mNPs (black) and NDs (grey). TEM images offer a more precise determination of particle and complex sizes than DLS measurements. The size of mNPs is roughly 20 nm, consistent with the nominal size, and they exhibit a spherical shape. In contrast, the size of individual NDs is more variable, ranging from particles under 20 nm (particularly observed in OND samples) to up to 100 nm. Additionally, NDs exhibit a broad variety of shapes, rarely spherical. Notably, agglomeration was detected only for nanodiamonds, whereas the mNPs remained well-dispersed with distinguishable boundaries. Nanocomplexes formed in all samples without significant variation based on ND termination, indicating that the termination did not substantially affect the assembly process as previously suggested by spectrophotometry experiments (Figure 5). During imaging, the least complexes were observed in the AgHND25 and AuOND25 samples. Most nanocomplexes contained 1 to 3 mNPs on the ND surface, with a clear distribution, which correlates with the observed stable plasmonic wavelength. However, the thin gaps between mNPs and NDs, predicted by our simulations, could not be resolved experimentally due to limitation in the imaging resolution.

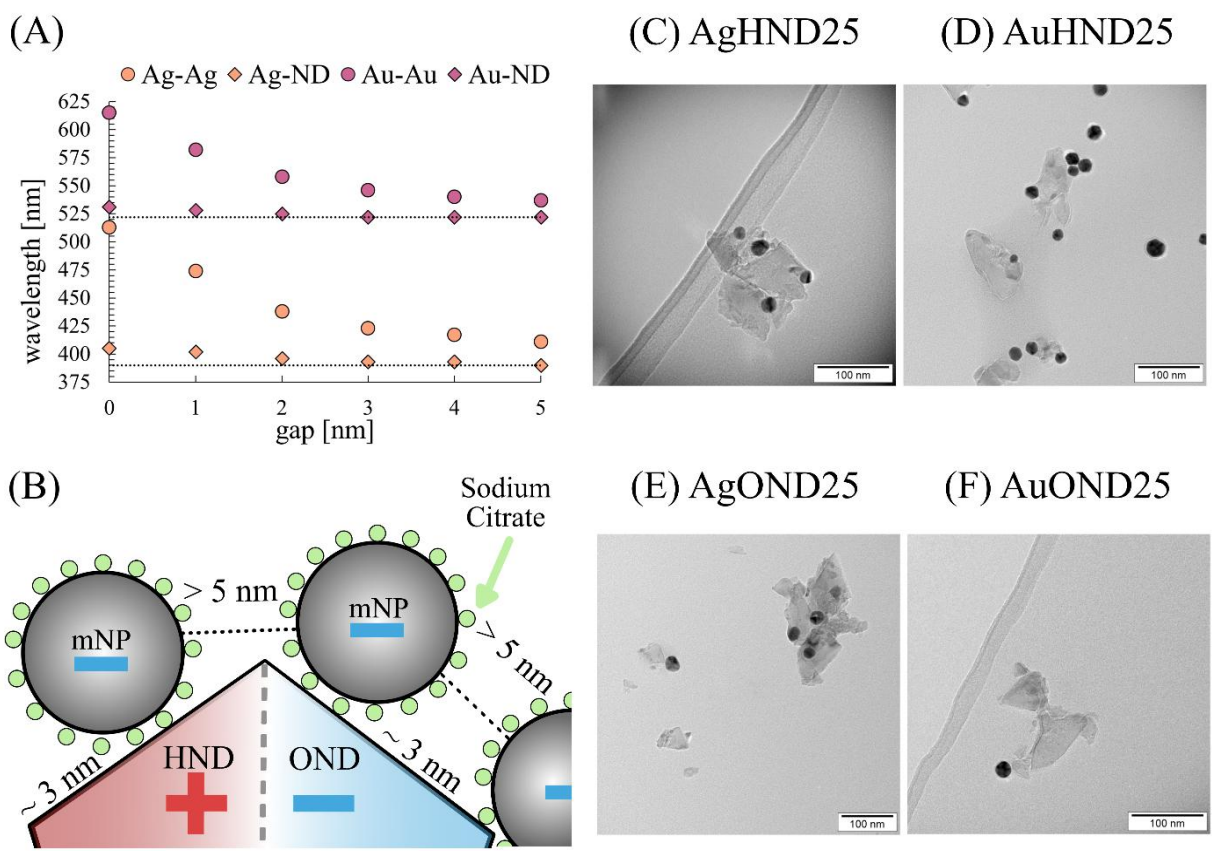

*Figure 7:* (A) Plasmonic absorption peak position of mNP-ND and mNP-mNP dimers in water medium as evaluated from the electromagnetic field simulations using COMSOL Multiphysics RF Module. The distance between the particles is varied from 5 nm to direct contact (0 nm). The dashed lines indicate the plasmonic peak position in pristine AgNP or AuNP colloid. (B) Geometrical scheme of the mNP-ND nanocomplexes indicating the minimum necessary gaps to prevent plasmonic wavelength shift. TEM images of colloidal mixtures - (C) AgHND25, (D) AuHND25, (E) AgOND25, and (F) AuOND25. The nanoparticle concentrations in these mixtures were ND 25 µg/mL, AgNP 9 µg/mL and AuNP 22.5 µg/mL.

3.5. *Plasmonic Model*

The agreement between the simulations and experiments may help answer the question of what causes the observed plasmonic behaviour of the nanocomplexes: at lower concentrations of mNPs in the nanocomplex, the plasmonic absorption peak increases, and at higher concentrations it decreases below the value of the mNP alone. The trend is consistent across all types of colloidal mixtures and reproducible across all experiments. At the same time, the plasmonic peak wavelength, i.e., the plasmonic frequency, does not change. Therefore, we can exclude aggregation of mNPs and generally all close plasmonic coupling as the cause of the changes. This is clearly supported by theory [41] as well as the simulations and experimental data presented here.

To explain this behaviour, we propose that two opposing mechanisms are acting simultaneously: charge redistribution, which enhances plasmonic absorption and is dominant at lower mNP concentrations, and plasmonic interference, which suppresses absorption and becomes the leading mechanism at higher mNP concentrations. We base this hypothesis on several arguments explained below.

Generally, the absorbance could also be decreased by Mie scattering (absorbance in transmitted light is affected both by absorption and scattering). However, since all our samples are optically highly transparent (Figure 1(B)), this effect is unlikely. Also note that the plasmonic intensity changes are evaluated after subtracting the pristine nanodiamond absorbance spectra.

Colour centres in nanodiamonds did not manifest themselves in the ND absorption spectra (Figure 1 (C)) nor in the spectra of colloidal mixtures (Figure 2), meaning they likely do not contribute to the observed plasmonic effects.

Another mechanism could be direct or indirect charge transfer, [5] a broadly explored effect with many diverse applications. [42] The possibility of charge transfer from AgNPs to nanodiamonds was previously proposed in the study of AgNP/ND/$C_3N_4$ heterostructures with enhanced visible-light photocatalytic performance. [43] The transferred free electrons can also be detected by infrared spectroscopy as has been shown for nanodiamonds [24], [44] or AuNR-$TiO_2$ core-shell structures. [5] However, for HPHT nanodiamonds, the electronic properties of which resemble bulk diamond [24], the energetic level alignment is not favourable for direct or indirect electron transfer from AgNPs (or AuNPs). The work function of AgNPs approaches 5.3 eV when AgNP size decreases below 30 nm, compared to typical bulk value of 4.3 eV. [45] It is well below the conduction band of diamond, in fact even below the valence band energy of 4.4 eV to 4.9 eV of hydrogenated diamond. [46]

Moreover, the direct charge transfer pathway provides an additional decay channel for plasmon damping, hence affecting the total scattering linewidth. [5] Nevertheless, in the absorption spectra we did not observe any significant changes in plasmonic peak linewidth. Thus, direct charge transfer is likely not the mechanism at work here. Indirect charge transfer to diamond would remain invisible in the plasmonic spectrum (and its intensity) because it occurs after plasmon decay. [5]

Based on the energy level alignment of mNPs and NDs, electron transfer could occur in the opposite direction (from NDs to mNPs) with the nanodiamond acting as sensitizer for higher energy wavelengths. [47] More electrons participating in the oscillation could cause the enhancement of plasmonic resonance, however, the plasmonic wavelength would shift with an increasing number of free electrons approximately as $\approx 1/\sqrt{n}$ [4], causing a blue shift, which was not observed in the experiments or simulations. One could assume that the number of transferred electrons is too low to cause a noticeable shift. However, then they could hardly cause a 120% increase in the plasmonic peak intensity. Note also that the 3 nm gap between mNP and ND reduces the probability of any charge transfer, though a plasmonic energy transfer (PRET) [48] could occur easily across such distance.

Energy transfer occurs from a higher excitation energy donor to a lower excitation energy acceptor or more precisely, depending on their spectral overlap. Since NDs absorb at higher energy than mNPs, they could capture the high energy part of the spectra and provide excitation energy to plasmonic particles in sufficient vicinity (plasmonic energy transfer can occur up to roughly 20 nm in distance). Such localized non-radiative energy transfer from a dielectric to a metal causing stimulated plasmon amplification (spaser) has already been reported. [49] Although the energy transfer from diamonds to mNPs could explain the plasmonic enhancement, it cannot explain the suppression, for which the energy transfer would have to occur in the opposite direction. Moreover, PRET manifests by quenching dips in the donor absorption (or scattering) spectra [48], which we did not observe in our experiments (compare Figure 1 (C) and Figure 2).

Related to the energy transfer is the Purcell effect, an interference effect, where the oscillator (plasmonic particle) radiates the wave which is reflected from the environment. In turn, the reflection excites the oscillator in or out of phase, leading to oscillator enhancement or suppression. Plasmonic resonance depends in general on the refractive index of the surroundings [1], [6], [7], [50], which could be altered by NDs. The compensation of the loss in a plasmonic metal by the gain in nearby dielectric has previously been demonstrated in a mixture of silver nanoparticles and rhodamine. [51] However, the concentration of NDs was constant, the refractive index had the same effect on the metal nanoparticles across all mNP concentrations.

The remaining mechanism for the plasmonic enhancement in mNP-ND nanocomplexes is the simplest one as already hinted by our simulations: charge redistribution and resulting field focusing between the mNPs and NDs. Such a local radiative density of optical states leading to dipole or quadrupole charge redistribution at plasmonic nanocavities has been observed between Au tips and silver substrates in scanning tunnelling microscopy. [52] In our case both types of NDs can mediate the local distribution of mNP electrons (as shown in Figure 1 (F) and 6 (D)) without altering the global electron concentration. As HPHT HNDs exhibit conductivity [24] they can readily contribute to the dipole across the gap with mNPs. Although ONDs are highly electrically resistive, their surface is polarized due to oxygen-related chemistry (negative zeta potential) and they are prone to trapping charge carriers, typically holes, in surface states in a rather high density. [53]

However, neither of the mechanisms discussed above can explain the suppression of the plasmonic resonance below the values of pristine mNPs. Hence there must be another counteracting mechanism that decreases the plasmonic resonance with an increasing mNP concentration. The suppression showed itself in the spectra without a wavelength shift at higher concentrations of mNPs.

As we compare relative differences of absorption between pristine mNP and ND-mNP colloids with the same mNP concentration, the suppressed LSPR absorption of ND-mNP complexes versus pristine colloids cannot be due to mere optical absorption on the particles around NDs. Actual plasmon resonance quenching would typically cause a broader spectral response due to the increased damping of plasmon oscillations, but our results show a stable peak width. Resonance quenching would also manifest itself by a plasmonic peak shift, which was not detected during our experiment either. Furthermore, the computational model showed that metal nanoparticles close to each other ($\leq 5$ nm) would cause a red-shift of plasmonic peak (Figure 7 (A)).

Thus, the plasmonic suppression at higher mNPs concentrations is attributed to interference effects. For the increasing number of mNPs accumulated around ND, the redistribution of electromagnetic field density takes place as shown by the simulations in Figure 1 (F) and Figure 6 (D). Although the maximum electromagnetic field is 4 times higher (41 V/m vs 11 V/m) in favour of the cluster in Figure 6 (D), an uneven distribution and proximity of particles can cause destructive interference of electromagnetic fields of adjacent mNPs [7] and thereby lead to the suppression of plasmonic absorption even below that of mNPs alone.

This effect is obvious from simulations of mNP nanoclusters without NDs in Figure 6 (C) where the saturation and decrease of absolute plasmonic peak intensity is observed despite increasing mNP numbers, i.e. concentration, in the nanocluster, for which one would simply expect further increase in absorption due to more particles present.

In this respect, one can also notice in the experiments in Figure 5 that samples AuHND25 with 45 µg/mL AuNP (100%), and AuHND6 with 11.25 µg/mL AuNP (25%) have a similar mixing concentration ratio of 5:9 (or 1:5 in number concentrations), yet the plasmonic enhancement effect of nanodiamond is positive for the latter, more diluted sample. Similar results may be seen for AgOND25 with 18 µg/mL AgNP (100%) and AgOND6 with 4.5 µg/mL AgNP (25%) having a mixing ratio of 4:3 (or 1:4 in number concentrations). This may indicate that in more diluted solutions, the spontaneous nanocomposite assembly is less efficient and thus reduces the interference effects in mNPs around NDs, which suppress the plasmonic enhancement.

Based on the results and discussion above, **Figure 8** schematically depicts the general plasmonic effect observed in the mNP-ND nanocomplexes and the proposed plasmonic model of two competing mechanisms, which explain the enhancement and suppression of the plasmonic absorption depending on mNP concentrations. In brief, the scheme shows the transition between the two opposing mechanisms – enhancement via charge redistribution and field focusing in the mNP-ND gap (increase of electron density on the surface of mNPs complemented by holes/polarization in NDs) and suppression via electromagnetic interference among the mNPs. The suppression caused by destructive interference of electromagnetic fields in mNPs comes into effect with more metal nanoparticles assembled around NDs.

These interactions between mNPs and NDs can also influence other optical properties of particles, such as photoluminescence (PL). To explore this relationship, we conducted PL measurements. The resulting spectra are provided in the Supplementary material (Figure S2). The PL results are inconclusive so far, suggesting the need for further investigation to obtain a deeper understanding of the connection between changes in plasmonic resonance intensity arising from the formation of mNP-ND complexes and photoluminescence, and other NP properties.

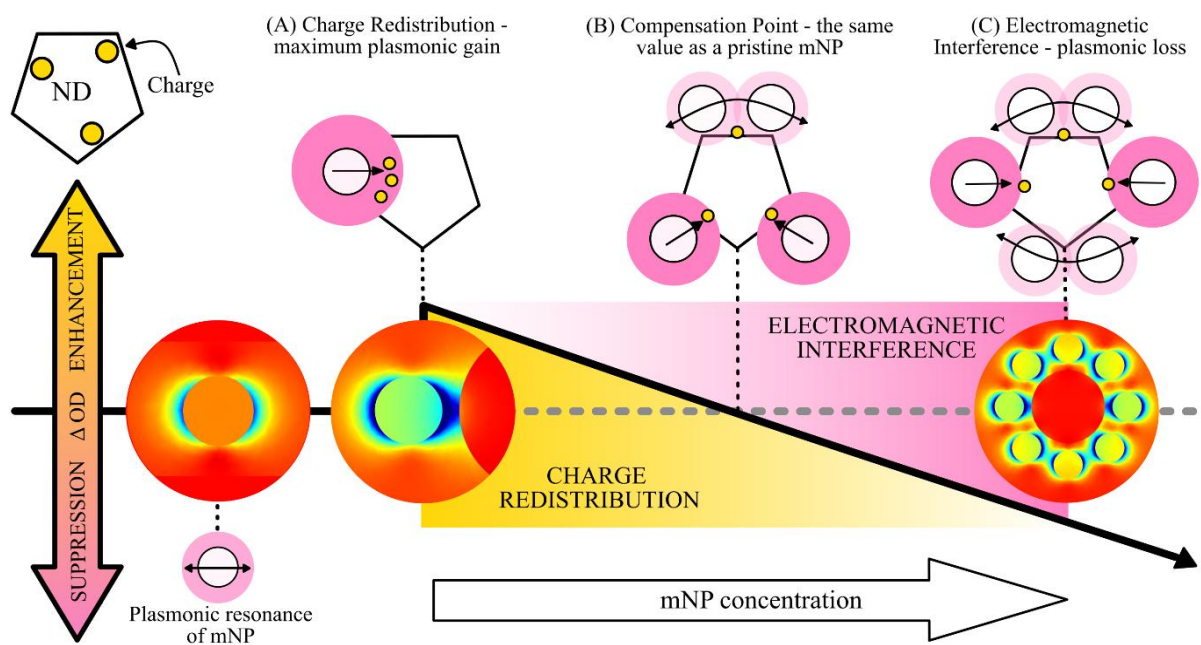

*Figure 8: Schematic model of two competing interactions that affect plasmonic resonance of mNP-ND nanocomplex: (A) enhancement via charge redistribution and field focusing, (B) compensation point where both interactions cancel out, and (C) suppression via electromagnetic interference of mNP fields. Arrows on nanoparticles denote plasmonic vibrations, their focusing or interference.*

## 4. Conclusion

This work investigated formation and properties of plasmonically active nanocomplexes from colloidal mixtures of metal nanoparticles (AgNPs and AuNPs) and surface modified HPHT nanodiamonds. An advantage of using the colloidal mixture approach to form nanocomplexes is in the easiness of preparation and well-defined size distribution, shape, surface chemistry, and concentration of employed particles. Surprisingly, the nanocomplexes formed in all colloidal mixtures spontaneously and independently of the ND zeta potential positive, or negative polarity, and mNP type.

The presence of NDs affected the intensity of plasmonic resonance in all cases. In the case of lower concentrations of mNPs, we observed relative enhancement of plasmonic resonance intensity in optical absorption spectra, whereas at higher concentrations there was a relative suppression of intensity to even below the values of pristine mNPs. The trend remained the same for all ND and mNP types and was consistent with electromagnetic field simulations. We observed the greatest enhancement in colloidal mixtures with low concentrations of AgNPs, where the plasmonic gain was more than double the original values. Altogether, NDs affected the plasmonic resonance of silver nanoparticles more than that of gold nanoparticles, stemming from the generally stronger plasmonic effect of silver.

With the plasmonic effect enhancement or suppression, the plasmonic frequency remained stable, which may be a useful feature for sensing applications. This means that the metal nanoparticles were not aggregated on nanodiamond and still acted as standalone units. This assertion was corroborated by electromagnetic field simulations. The simulations also showed that for a stable plasmonic frequency the metal nanoparticles must also have a nm-thin gap to nanodiamonds. The gap can be explained by the citrate passivation layer on metal nanoparticles. We proposed that citrate is also the cause of spontaneous composite formation on both HNDs and ONDs, in spite of their opposite zeta potential.

The plasmonic frequency did not change even at lower concentrations of NDs in the mixtures and such mixtures exhibited the same plasmonic enhancement trend depending on mNPs concentration. The mNP-ND complex colloidal assembly thus seems inherently self-controlled and the arising plasmonic effects are robust.

Based on the experiments and simulations we proposed a model where two opposing interactions are in effect simultaneously – charge redistribution and field focusing enhancing the plasmonic effect, and collective electromagnetic interference suppressing it. We believe that this study will shed more light on the properties and effects of nanocomplexes of NDs with plasmonic particles and thereby helps elucidate some prior observations in literature as well as open prospects for further applications from sensing and diamond quantum centres control to catalysis and energy conversion. [42]

**Acknowledgement**

This work was supported by the Ministry of Education, Youth and Sports of the Czech Republic project CZ.02.01.01/00/22_008/0004596 (SenDISo) under the Operational Program Johannes Amos Comenius in the call Excellent Research, and CzechNanoLab Research Infrastructure (LM2023051). The work was also supported by the Czech Science Foundation project no. 23-04322L, and National Science Center project no. 2021/43/I/ST7/03205 under the OPUS call in the Weave program. Student support from project SGS23/166/OHK4/3T/13 is gratefully appreciated. We kindly acknowledge fruitful discussions with Robert Bogdanowicz on metal nanoparticles interactions with diamond and carbon nanomaterials.

**CRediT AUTHORSHIP CONTRIBUTION STATEMENT**

**Vendula Hrnčířová:** Data curation, Formal analysis, Investigation, Visualization, Writing – original draft, **Markéta Šlapal Bařinková:** Formal analysis, Investigation, Methodology, Project Administration, Software, Writing - review & editing, **Muhammad Qamar:** Data curation, Formal analysis, Investigation, Methodology, **Kateřina Kolářová:** Investigation, Resources, **Jaroslav Kuliček:** Investigation, **Štěpán Stehlík:** Investigation, Writing - review & editing, **Alexander Kromka:** Resources, Funding acquisition, Writing – review & editing, **Bohuslav Rezek:** Conceptualization, Formal analysis, Investigation, Funding acquisition, Project Administration, Resources, Supervision, Writing - review & editing

**OPEN DATA DOI:** [10.5281/zenodo.14512705](10.5281/zenodo.14512705)

**REFERENCES**

[1] Y.-S. Borghei, S. Hosseinkhani, and M. R. Ganjali, '"Plasmonic Nanomaterials": An emerging avenue in biomedical and biomedical engineering opportunities', *Journal of Advanced Research*, vol. 39, pp. 61–71, Jul. 2022, doi: 10.1016/j.jare.2021.11.006.

[2] C. Liu *et al.*, 'Recent advances of plasmonic nanofluids in solar harvesting and energy storage', *Journal of Energy Storage*, vol. 72, p. 108329, Nov. 2023, doi: 10.1016/j.est.2023.108329.

[3]     A. Gellé and A. Moores, 'Plasmonic nanoparticles: Photocatalysts with a bright future', *Current Opinion in Green and Sustainable Chemistry*, vol. 15, pp. 60–66, Feb. 2019, doi: 10.1016/j.cogsc.2018.10.002.

[4]     N. Zhang, C. Han, X. Fu, and Y.-J. Xu, 'Function-Oriented Engineering of Metal-Based Nanohybrids for Photoredox Catalysis: Exerting Plasmonic Effect and Beyond', *Chem*, vol. 4, no. 8, pp. 1832–1861, Aug. 2018, doi: 10.1016/j.chempr.2018.05.005.

[5]     B. Ostovar *et al.*, 'The role of the plasmon in interfacial charge transfer', *Science Advances*, vol. 10, no. 27, p. eadp3353, Jul. 2024, doi: 10.1126/sciadv.adp3353.

[6]     E. Petryayeva and U. J. Krull, 'Localized surface plasmon resonance: Nanostructures, bioassays and biosensing—A review', *Analytica Chimica Acta*, vol. 706, no. 1, pp. 8–24, Nov. 2011, doi: 10.1016/j.aca.2011.08.020.

[7]     J. Liu *et al.*, 'Recent Advances of Plasmonic Nanoparticles and their Applications', *Materials (Basel)*, vol. 11, no. 10, p. 1833, Sep. 2018, doi: 10.3390/ma11101833.

[8]     A. Amirjani and D. F. Haghshenas, 'Ag nanostructures as the surface plasmon resonance (SPR)-based sensors: A mechanistic study with an emphasis on heavy metallic ions detection', *Sensors and Actuators B: Chemical*, vol. 273, pp. 1768–1779, Nov. 2018, doi: 10.1016/j.snb.2018.07.089.

[9]     S. Orlanducci, 'Gold-Decorated Nanodiamonds: Powerful Multifunctional Materials for Sensing, Imaging, Diagnostics, and Therapy', *European Journal of Inorganic Chemistry*, vol. 2018, no. 48, pp. 5138–5145, 2018, doi: 10.1002/ejic.201800793.

[10]    Z. Wang, F. Zhang, A. Ning, D. Lv, G. Jiang, and A. Song, 'Nanosilver supported on inert nano-diamond as a direct plasmonic photocatalyst for degradation of methyl blue', *Journal of Environmental Chemical Engineering*, vol. 9, no. 1, p. 104912, Feb. 2021, doi: 10.1016/j.jece.2020.104912.

[11]    K. Kolwas, 'Decay Dynamics of Localized Surface Plasmons: Damping of Coherences and Populations of the Oscillatory Plasmon Modes', *Plasmonics*, vol. 14, no. 6, pp. 1629–1637, Dec. 2019, doi: 10.1007/s11468-019-00958-1.

[12]    S. Iqbal, M. S. Rafique, N. Iqbal, S. Akhtar, A. A. Anjum, and M. B. Malarvili, 'Synergistic effect of Silver-Nanodiamond composite as an efficient antibacterial agent against *E. coli* and *S. aureus*', *Heliyon*, vol. 10, no. 9, p. e30500, May 2024, doi: 10.1016/j.heliyon.2024.e30500.

[13]    D. Lee, S. H. Jeong, and E. Kang, 'Nanodiamond/gold nanorod nanocomposites with tunable light-absorptive and local plasmonic properties', *Journal of Industrial and Engineering Chemistry*, vol. 65, pp. 205–212, Sep. 2018, doi: 10.1016/j.jiec.2018.04.030.

[14]    Y.-C. Lin *et al.*, 'Multimodal bioimaging using nanodiamond and gold hybrid nanoparticles', *Sci Rep*, vol. 12, no. 1, p. 5331, Mar. 2022, doi: 10.1038/s41598-022-09317-3.

[15]    L. Schmidheini, R. F. Tiefenauer, V. Gatterdam, A. Frutiger, T. Sannomiya, and M. Aramesh, 'Self-Assembly of Nanodiamonds and Plasmonic Nanoparticles for Nanoscopy', *Biosensors*, vol. 12, no. 3, p. 148, Feb. 2022, doi: 10.3390/bios12030148.

[16]    L. Liang, P. Zheng, S. Jia, K. Ray, Y. Chen, and I. Barman, 'Plasmonic Nanodiamonds', *Nano Lett.*, vol. 23, no. 12, pp. 5746–5754, Jun. 2023, doi: 10.1021/acs.nanolett.3c01514.

[17]    B.-M. Chang, L. Pan, H.-H. Lin, and H.-C. Chang, 'Nanodiamond-supported silver nanoparticles as potent and safe antibacterial agents', *Sci Rep*, vol. 9, no. 1, p. 13164, Sep. 2019, doi: 10.1038/s41598-019-49675-z.

[18]    P.-C. Tsai, C. P. Epperla, J.-S. Huang, O. Y. Chen, C.-C. Wu, and H.-C. Chang, 'Measuring Nanoscale Thermostability of Cell Membranes with Single Gold–Diamond Nanohybrids', *Angewandte Chemie International Edition*, vol. 56, no. 11, pp. 3025–3030, 2017, doi: 10.1002/anie.201700357.

[19]    J.-X. Qin *et al.*, 'Nanodiamonds: Synthesis, properties, and applications in nanomedicine', *Materials & Design*, vol. 210, p. 110091, Nov. 2021, doi: 10.1016/j.matdes.2021.110091.

[20]    J. Xu and E. K.-H. Chow, 'Biomedical applications of nanodiamonds: From drug-delivery to diagnostics', *SLAS Technology*, vol. 28, no. 4, pp. 214–222, Aug. 2023, doi: 10.1016/j.slast.2023.03.007.

[21]    F. Pan *et al.*, 'Recent advances in the structure and biomedical applications of nanodiamonds and their future perspectives', *Materials & Design*, vol. 233, p. 112179, Sep. 2023, doi: 10.1016/j.matdes.2023.112179.

[22]    V. N. Mochalin, O. Shenderova, D. Ho, and Y. Gogotsi, 'The properties and applications of nanodiamonds', *Nature Nanotech*, vol. 7, no. 1, pp. 11–23, Jan. 2012, doi: 10.1038/nnano.2011.209.

[23]    D. Miliaieva *et al.*, 'Absolute energy levels in nanodiamonds of different origins and surface chemistries', *Nanoscale Adv.*, vol. 5, no. 17, pp. 4402–4414, Aug. 2023, doi: 10.1039/D3NA00205E.

[24]    S. Stehlik *et al.*, 'Electrical and colloidal properties of hydrogenated nanodiamonds: Effects of structure, composition and size', *Carbon Trends*, vol. 14, p. 100327, Mar. 2024, doi: 10.1016/j.cartre.2024.100327.

[25]    S. Stehlik *et al.*, 'Size and Purity Control of HPHT Nanodiamonds down to 1 nm', *J. Phys. Chem. C*, vol. 119, no. 49, pp. 27708–27720, Dec. 2015, doi: 10.1021/acs.jpcc.5b05259.

[26]    K. Kolarova, D. Miliaieva, and S. Stehlik, 'Polyvinylpyrrolidone coating for nanodiamond stabilization in saline solution and silver nanoparticle decoration', presented at the NANOCON 2020, 2020, pp. 80–85. doi: 10.37904/nanocon.2020.3696.

[27]    B. Woodhams *et al.*, 'Graphitic and oxidised high pressure high temperature (HPHT) nanodiamonds induce differential biological responses in breast cancer cell lines', *Nanoscale*, vol. 10, no. 25, pp. 12169–12179, Jul. 2018, doi: 10.1039/C8NR02177E.

[28]    K. Kolarova *et al.*, 'Hydrogenation of HPHT nanodiamonds and their nanoscale interaction with chitosan', *Diamond and Related Materials*, vol. 134, p. 109754, Apr. 2023, doi: 10.1016/j.diamond.2023.109754.

[29]   P. Aprà *et al.*, 'Interaction of Nanodiamonds with Water: Impact of Surface Chemistry on Hydrophilicity, Aggregation and Electrical Properties', *Nanomaterials*, vol. 11, no. 10, Art. no. 10, Oct. 2021, doi: 10.3390/nano11102740.

[30]   *MATLAB*. (2023). The MathWorks, Inc., Natick, Massachusetts.

[31]   D. Nečas and P. Klapetek, 'Gwyddion: an open-source software for SPM data analysis', *centr.eur.j.phys.*, vol. 10, no. 1, pp. 181–188, Feb. 2012, doi: 10.2478/s11534-011-0096-2.

[32]   'Refractive index of CRYSTALS - diamond'. Accessed: Dec. 10, 2024. [Online]. Available: https://refractiveindex.info/?shelf=3d&book=crystals&page=diamond

[33]   'A Table of Electrical Conductivity and Resistivity of Common Materials', ThoughtCo. Accessed: Dec. 10, 2024. [Online]. Available: https://www.thoughtco.com/table-of-electrical-resistivity-conductivity-608499

[34]   P. B. Johnson and R. W. Christy, 'Optical Constants of the Noble Metals', *Phys. Rev. B*, vol. 6, no. 12, pp. 4370–4379, Dec. 1972, doi: 10.1103/PhysRevB.6.4370.

[35]   C. Marchal, L. Saoudi, H. A. Girard, V. Keller, and J.-C. Arnault, 'Oxidized Detonation Nanodiamonds Act as an Efficient Metal-Free Photocatalyst to Produce Hydrogen Under Solar Irradiation', *Advanced Energy and Sustainability Research*, vol. 5, no. 3, p. 2300260, 2024, doi: 10.1002/aesr.202300260.

[36]   T. Hainschwang, F. Notari, and G. Pamies, 'The origin of color of 1330 nm center diamonds', *Diamond and Related Materials*, vol. 110, p. 108151, Dec. 2020, doi: 10.1016/j.diamond.2020.108151.

[37]   I. Machova *et al.*, 'The bio-chemically selective interaction of hydrogenated and oxidized ultra-small nanodiamonds with proteins and cells', *Carbon*, vol. 162, pp. 650–661, Jun. 2020, doi: 10.1016/j.carbon.2020.02.061.

[38]   J. Jira *et al.*, 'Inhibition of E. coli Growth by Nanodiamond and Graphene Oxide Enhanced by Luria-Bertani Medium', *Nanomaterials*, vol. 8, no. 3, Art. no. 3, Mar. 2018, doi: 10.3390/nano8030140.

[39]   I. Dudeja, R. K. Mankoo, A. Singh, and J. Kaur, 'Citric acid: An ecofriendly cross-linker for the production of functional biopolymeric materials', *Sustainable Chemistry and Pharmacy*, vol. 36, p. 101307, Dec. 2023, doi: 10.1016/j.scp.2023.101307.

[40]   L. K. Sørensen *et al.*, 'Nature of the Anomalous Size Dependence of Resonance Red Shifts in Ultrafine Plasmonic Nanoparticles', *J. Phys. Chem. C*, vol. 126, no. 39, pp. 16804–16814, Oct. 2022, doi: 10.1021/acs.jpcc.2c03738.

[41]   P. K. Jain, W. Huang, and M. A. El-Sayed, 'On the Universal Scaling Behavior of the Distance Decay of Plasmon Coupling in Metal Nanoparticle Pairs: A Plasmon Ruler Equation', *Nano Lett.*, vol. 7, no. 7, pp. 2080–2088, Jul. 2007, doi: 10.1021/nl071008a.

[42]   H. Tang *et al.*, 'Plasmonic hot electrons for sensing, photodetection, and solar energy applications: A perspective', *The Journal of Chemical Physics*, vol. 152, no. 22, p. 220901, Jun. 2020, doi: 10.1063/5.0005334.

[43]   L.-X. Su, Q. Lou, C.-X. Shan, D.-L. Chen, J.-H. Zang, and L.-J. Liu, 'Ag/Nanodiamond/g-C3N4 heterostructures with enhanced visible-light photocatalytic performance', *Applied Surface Science*, vol. 525, p. 146576, Sep. 2020, doi: 10.1016/j.apsusc.2020.146576.

[44]   O. S. Kudryavtsev *et al.*, 'Fano-type Effect in Hydrogen-Terminated Pure Nanodiamond', *Nano Lett.*, vol. 22, no. 7, pp. 2589–2594, Apr. 2022, doi: 10.1021/acs.nanolett.1c04887.

[45]   M. Schnippering, M. Carrara, A. Foelske, R. Kötz, and D. J. Fermín, 'Electronic properties of Ag nanoparticle arrays. A Kelvin probe and high resolution XPS study', *Phys. Chem. Chem. Phys.*, vol. 9, no. 6, pp. 725–730, Feb. 2007, doi: 10.1039/B611496B.

[46]   B. Rezek and C. E. Nebel, 'Kelvin force microscopy on diamond surfaces and devices', *Diamond and Related Materials*, vol. 14, no. 3, pp. 466–469, Mar. 2005, doi: 10.1016/j.diamond.2005.01.041.

[47]   J. Henych *et al.*, 'Reactive adsorption and photodegradation of soman and dimethyl methylphosphonate on TiO2/nanodiamond composites', *Applied Catalysis B: Environmental*, vol. 259, p. 118097, Dec. 2019, doi: 10.1016/j.apcatb.2019.118097.

[48]   Y. Choi, T. Kang, and L. P. Lee, 'Plasmon Resonance Energy Transfer (PRET)-based Molecular Imaging of Cytochrome c in Living Cells', *Nano Lett.*, vol. 9, no. 1, pp. 85–90, Jan. 2009, doi: 10.1021/nl802511z.

[49]   D. J. Bergman and M. I. Stockman, 'Surface Plasmon Amplification by Stimulated Emission of Radiation: Quantum Generation of Coherent Surface Plasmons in Nanosystems', *Phys. Rev. Lett.*, vol. 90, no. 2, p. 027402, Jan. 2003, doi: 10.1103/PhysRevLett.90.027402.

[50]   J. J. Mock, D. R. Smith, and S. Schultz, 'Local Refractive Index Dependence of Plasmon Resonance Spectra from Individual Nanoparticles', *Nano Lett.*, vol. 3, no. 4, pp. 485–491, Apr. 2003, doi: 10.1021/nl0340475.

[51]   M. A. Noginov *et al.*, 'The effect of gain and absorption on surface plasmons in metal nanoparticles', *Appl. Phys. B*, vol. 86, no. 3, pp. 455–460, Feb. 2007, doi: 10.1007/s00340-006-2401-0.

[52]   A. Martín-Jiménez *et al.*, 'Unveiling the radiative local density of optical states of a plasmonic nanocavity by STM', *Nat Commun*, vol. 11, no. 1, p. 1021, Feb. 2020, doi: 10.1038/s41467-020-14827-7.

[53]   Y. Itoh, Y. Sumikawa, H. Umezawa, and H. Kawarada, 'Trapping mechanism on oxygen-terminated diamond surfaces', *Applied Physics Letters*, vol. 89, no. 20, p. 203503, Nov. 2006, doi: 10.1063/1.2387983.